# Secure quantum imaging with decoy state heralded single photons

SIDDHANT VERNEKAR[1,*] AND JOLLY XAVIER[1,*]

[1]SeNSE, Indian Institute of Technology Delhi, Hauz Khas, New Delhi 110016, India
*Corresponding author: idz20228534@iitd.ac.in , jxavier@sense.iitd.ac.in

*Weak coherent source (WCS) and spontaneous parametric down-converted heralded single photon pairs have found applications in quantum key distribution (QKD) and quantum imaging (QI) experiments. Decoy state methods have also been used to enhance the security for QKD and QI. We study quantum-secured imaging with the decoy-state heralded single photon source (HSPS). The HSPS's superior performance in low-photon-number regimes makes it an ideal candidate for integrating quantum key distribution protocols to reduce measurement uncertainty and ensure secure QI. Furthermore, our results also infer the influence of the decoy state WCS, due to its higher operating speed than decoy state HSPS, would be effective in conditions that allow higher mean photon numbers for quantum-secured imaging.*

Ubiquitous quantum properties of nonclassical light sources and their possibility of being integrated into systems for advanced sensing, imaging and secured information processing have been the focus of research and technological revolution in the recent past [1–3]. Weak coherent source (WCS) and spontaneous parametric down conversion (SPDC) based heralded single-photon sources (HSPS) have found applications in quantum key distribution(QKD) [4–6] and quantum imaging (QI) experiments [7–10]. The Decoy state method is helpful in practical situations when the source is not an ideal single-photon source but could also be embedded with multiple photon components in the transmitted information [11]. It is a technique used in QKD to improve the security of information transmitted. The secure key rate, $R$ is given by [11]:

$$R \geq q[Q_1[1 - H_2(e_1)] - Q_\mu f(E_\mu)H_2(E_\mu)]$$

Where $q$ is a parameter depending on the QKD protocol, $Q_1$ is the gain of single photons, $H$ is the binary Shannon entropy, $e_1$ is the error rate od single photon states, $E_\mu$ is the overall quantum bit error rate, and $f(E_\mu)$ is the error correction efficiency. This calculation ensures that a secure key is generated, accounting for the presence of multi-photon states. The decoy state method has been applied with WCS for QKD and QI experiments and also with HSPS for QKD implementation [12-14].

Correlated photon pairs from SPDC significantly enhance the signal-to-noise ratio, especially in scenarios with low light levels [15], and have been the basis of several advanced QI techniques [16–18]. An essential advancement in this field came with the realization that immediate detection of the idler photon, followed by correlating it with the returning signal later can be highly effective. This enhancement stems from the fact that the correlated photons in SPDC are generated simultaneously and maintain a time correlation [19]. A single-photon imaging experiment based on an SPDC source [10] has been recently implemented. A polarization-time correlation-based quantum-secured imaging (QSI) approach has been recently proposed [20] as an alternative to QKD-based imaging methods to counter intercept and resend attacks [13]. However, in the experimental realization [13], the Digital Micromirror Device (DMD) introduces global phases to the prepared quantum states. This can satisfy the phase randomization condition even without using a phase-randomized laser [21] and, when combined with a low average photon count can enhance security against such intercept-and-resend attacks for QKD-based imaging methods. Furthermore, the concept of quantum ghost polarimetry has been introduced [22 - 23]. This theory, relying on different polarization states, promises to enhance imaging quality significantly. Given these advancements, it is vital to compare single-photon source-based QKD protocols [12, 24–26], which can utilize various generated polarization states for improved imaging quality. Such comparisons are essential for secure imaging applications, highlighting the need for a detailed understanding of these different approaches and their implications.

We propose QSI with decoy-state HSPS and carries out a comparative analysis of WCS and HSPS with decoy states. We evaluate HSPS and WCS sources based on their statistical properties, similar to recent experimental demonstration [27], which have characterized their sources based on their statistical properties. The study is based on the assumption that achieving maximum accuracy and no compromise on the same is of utmost importance for sensitive imaging applications.

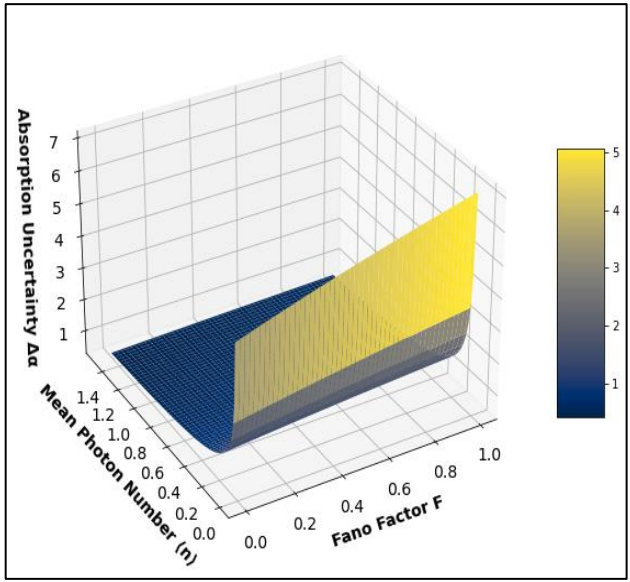

Fig. 1. The 3D plot of absorption uncertainty in relation to the Fano factor and mean photon number, here absorption factor $\alpha = 0.5$ is taken as a fixed value

Figure 1 illustrates a graphical representation of the relation among the uncertainty in absorption of a material, the Fano factor, and the mean photon number. The absorption uncertainty is crucial when dealing with the accuracy of measurements in biological samples or stealth imaging systems. Absorption uncertainty is given by the equation $\Delta\alpha = \sqrt{\frac{\alpha(1-\alpha)+F(1-\alpha)^2}{\langle\hat{n}\rangle}}$ ; $F$ is the Fano factor, $\langle\hat{n}\rangle$ is the mean photon number and $\alpha$ is the absorption factor [28]. When we have a lower mean photon number and if value of $F$ is less than one, like in the sub-Poissonian statistics photon source like HSPS, uncertainty is lower than higher $F$ values such as in the Poissonian source like WCS with $F$=1. The calculated value of $F$ estimated by [29], $F=\langle n\rangle[g^2(0)-1]+1$ for mean photon number 0.3 are 0.7015, 0.714 and 0.85 for $g^2(0)$= 0.005, 0.05 and 0.5, respectively. Similarly, the $F$ values for mean photon number 0.05 are 0.95025, 0.9525 and 0.975 for $g^2(0)$= = 0.005, 0.05 and 0.5, respectively. Furthermore, if the mean photon number can be increased to a favorable limit, it will also reduce the uncertainty in measurements with the WCS where $F$=1. This relationship is beneficial for applications that demand high levels of accuracy, such as biological imaging, where minimizing the exposure to light is crucial to avoid any potential harm to the samples. Stealth imaging, which aims to evade detection using low photon quantities, also benefits from a lower Fano factor.

The equation that describes the photon number distribution for a weak coherent source (Poissonian) is given by $P_{k,x}^{weak} = \frac{e^{-x} * x^k}{k!}$, and for a SPDC based HSPS(thermal) is given by [26], $p_{k,x}^{HSPS_{per}} = \frac{x^n}{(1+x)^{n+1}} * \frac{(1-(1-\eta_A)^k + d_A)}{P_x^{post}}$ where $x$ is the mean photon number, k is number of photons. $\eta_A$ represents the detection efficiency at the source end, $d_A$ is the dark count rate for Alice detector, $P_x^{post}$ is the post-selected probability. Figures 2a-b illustrate the comparison of single-photon contributions from two types of photon sources HSPS and WCS. It shows that in the area where the mean photon number is low, HSPS has a much greater chance of emitting a single photon. Additionally, HSPS exhibits sub-Poissonian statistics (variance less than the mean for photon numbers). Nevertheless, as the mean photon number rises closer to 0.6, the probability of single photons originating from both sources becomes similar. This is evident from the green curve (representing the heralded source) and the yellow dotted curve (representing the WCS) seen in Figure 2b. To guarantee security, the prevalence of single-photon events over multi-photon events is essential in QKD. The graph given in Figure 2 is a reference for determining optimal mean photon number of these sources. To minimize multi-photon occurrences, having a maximum mean photon number of around 0.2 for HSPS is preferable. Conversely, using WCS this numerical value may be increased to around 1.

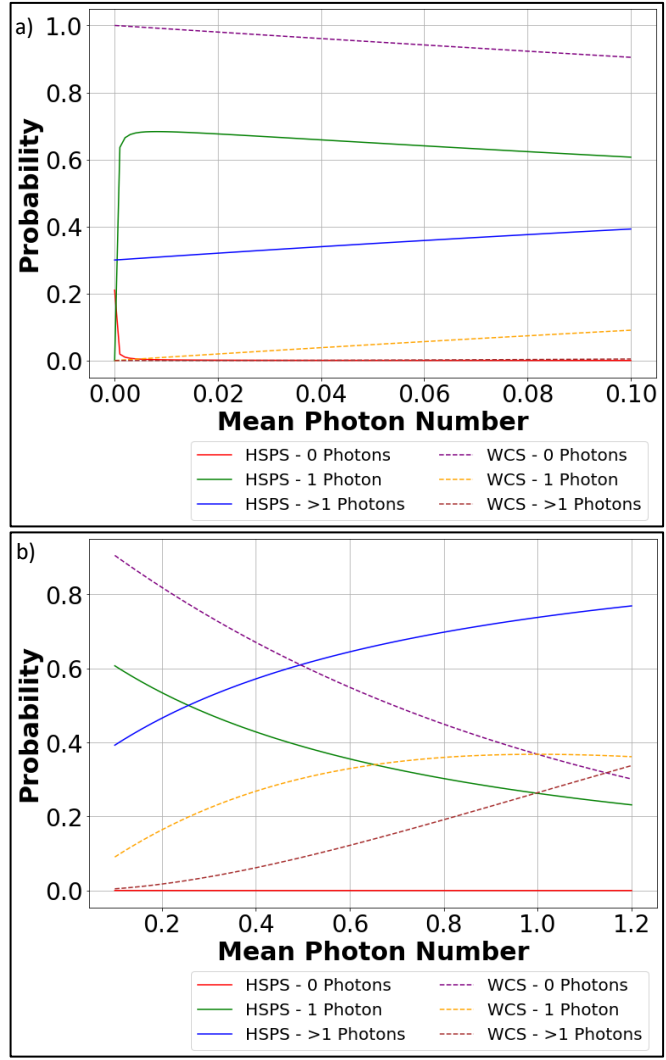

Fig. 2. Quantitative Analysis of Single-Photon Emissions from SPDC based Heralded Single Photon Source (HSPS) and Weak Coherent single (WCS) photon source: Implications for QSI. (correlation probability=0.7, $\eta_A = 0.5$ and, $d_A = 10^{-5}$) a) for mean photon number 0 to 0.1. b) for mean photon number>0.1 to 1.

Figure 3a shows plots illustrating the relation between the secure key rate and loss in a QKD system with WCS and HSPS, considering different mean photon numbers with $d_B = 10^{-6}$ as the dark count rate at the receiving detector. The figure displays a series of curves representing the different mean photon numbers of 0.01, 0.05, and 0.1 with combined decoys as 0.001 and 0. Each curve shows a noticeable decrease in the key rate as the distance increases. The figure 3a illustrates variations in the secure rate at lower mean photon numbers for WCS, indicating a less stable QSI performance across distances at these photon levels with WCS as compared to HSPS. Conversely, Figure 3b corresponds to higher mean photon numbers 0.2, 0.25, and 0.3 with combined decoys as 0.1 and 0. The curves are densely clustered and demonstrate a steady decrease for WCS. This suggests that the secure rate stays relatively constant across distances at higher mean photon numbers with less variability for WCS. This implies that it is advantageous to maintain a greater mean photon number for QSI with WCS since it leads to a more stable and reliable key rate than HSPS for the considered mean photon number regime. The comparison between the two plots highlights that lower mean photon numbers are more vulnerable to rate fluctuations for WCS than HSPS, which could impact the effectiveness of QSI with WCS. On the other hand, a larger mean photon number provides a more stable key rate, which can be advantageous for sustaining QSI across longer distances with WCS compared to HSPS.

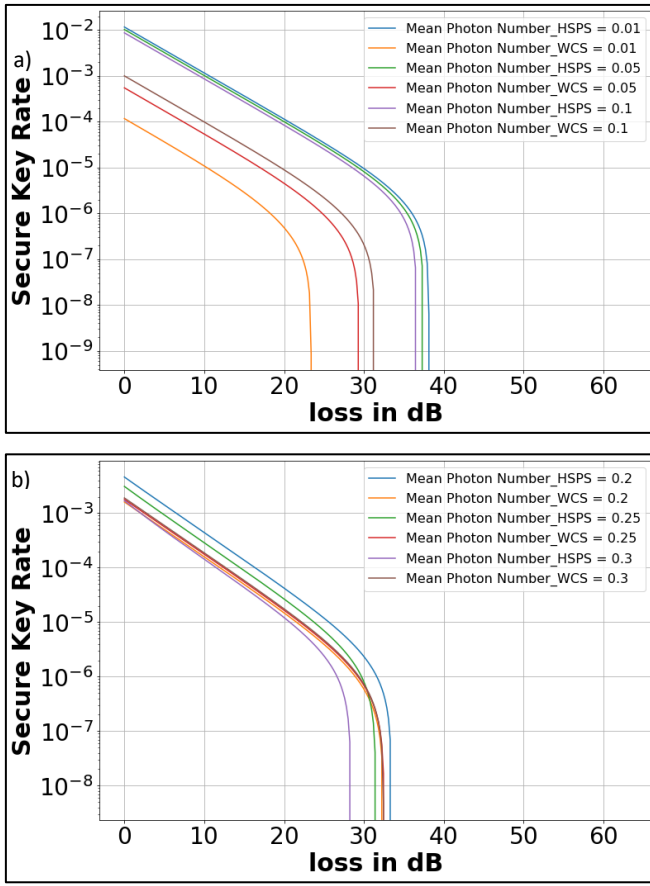

Fig. 3. Influence of mean photon number on the robustness of secure key generation rate in QSI with WCS and SPDC based single photon source at $d_B = 10^{-6}$ ,.a) For low mean photon numbers. b) For higher mean photon numbers.

We have assessed the widely prevalent two photon sources WCS and HSPS by considering their reliance on crucial factors such as the absorption uncertainty relation, distribution of photon numbers based on their statistics, and the secure key rate. Our technique involves a thorough analysis of the photon number distribution properties of both sources and the attainable secure key rate under QKD protocol implementation situations. HSPS decoy source (thermal photon number distribution statistics) display less fluctuation than WCS decoy in a low mean photon number region. On the other hand, WCS decoy (Poisssonian distribution statistics of photon number) would benefit in situations that need a larger mean photon number by implementing appropriate solutions against intercept and resend attacks [30,31]. The secure key rate for HSPS-decoy is higher than WCS-decoy at loss of 10 dB by an order of magnitude, and this advantage increases to 1.5 orders at greater loss (30dB), assuming optimal performance from both. However, the repetition rate for HSPS is in the MHz range [32]. In contrast, WCS operates in the GHz range [33], suggesting that WCS might generate more secure photons with current technology when the overall figure of merit is considered.

Moreover, the SPDC-based HSPS source is potentially valuable for biological imaging and stealth quantum secure satellite-based imaging applications due to its sub-Poissonian characteristics. HSPS also include a higher contribution of single photons and reduced fluctuations in secure key rate at low mean photon numbers with its thermal mean photon number statistics. Our case involves transmitting and detecting a signal after reflection from a target, resulting in a more significant loss. However, with a reconstruction algorithm, it is estimated that an image can be discernible with as few as 7000 photons [34]. The operating speed of spatial light modulator or DMD could further limit the overall system operation speed. Therefore, instead of using it at the source end, as in computational ghost imaging, using it at the receiver end might be more effective, similar to a single-pixel camera setup, to ensure meaningful photon counts from both sources when operating speed is a priority [10]. HSPS mean photon number statistics could also be changed to Poissonian from thermal by adjusting the pulsed laser duration with respect to the coherence time of the laser to reduce the multi photons contribution and match with the WCS statistics [35]. The emerging domain of QI has great potential for advancing biological and stealth imaging, particularly in scenarios where minimizing the photon exposure is essential to prevent harm and detection.

In conclusion, two types of single-photon sources SPDC as well as WCS have been investigated under low-light conditions. The SPDC-based HSPS decoy state source have the advantage of having sub-Poissonian statistics and quantum correlations to improve measurements and suppress background noise [9]. However, it is also to be noted that the SPDC photon generation rate is comparatively lower than the WCS-decoy version while considering the overall figure of merit with current technology. This in turn demands further research for new materials or techniques for the improvement in much more efficient SPDC photon pair generation. The present study has also significant implications for the QI domain since integrating the single photon sources with single-pixel imaging techniques is expected to improve the quality of images by using polarization states while also assuring quantum-secured imaging. Moreover, the presented evaluations shed light into right choice of single photon sources for quantum-enhanced imaging applications, especially in strategic or sensitive realms where security is a matter of great concern. Future investigations based on the current research would also involve the study of channel characteristics and the robustness of quantum imaging techniques to environmental conditions to quantify the details of imaging results based on the source parameters.

**Disclosures.** The authors declare no conflicts of interest.